\documentclass[prc,aps,twocolumn,preprintnumbers,nofootinbib,showkeys,showpacs,amsmath,amssymb,floatfix,superscriptaddress]{revtex4}
\usepackage[usenames]{color}
\usepackage{amsmath}
\usepackage{graphicx}
\usepackage{float,epsfig}
\usepackage{lscape}
\newcommand\be{\begin{equation}}
\newcommand\ee{\end{equation}}
\newcommand\bea{\begin{eqnarray}}
\newcommand\eea{\end{eqnarray}}

\usepackage{graphicx}
\usepackage[usenames]{color}
\usepackage{ulem}

\begin{document}

\title{From bare interactions, low--energy constants and unitary gas to 
nuclear density functionals without free parameters: application to neutron matter}
  
\author{Denis Lacroix} \email{lacroix@ipno.in2p3.fr}
\affiliation{Institut de Physique Nucl\'eaire, IN2P3-CNRS, Universit\'e Paris-Sud, Universit\'e Paris-Saclay, F-91406 Orsay Cedex, France}
  
\author{Antoine Boulet}
\affiliation{Institut de Physique Nucl\'eaire, IN2P3-CNRS, Universit\'e Paris-Sud, Universit\'e Paris-Saclay, F-91406 Orsay Cedex, France}
\author{Marcella Grasso}
\affiliation{Institut de Physique Nucl\'eaire, IN2P3-CNRS, Universit\'e Paris-Sud, Universit\'e Paris-Saclay, F-91406 Orsay Cedex, France}
 \author{C.-J. Yang}
\affiliation{Institut de Physique Nucl\'eaire, IN2P3-CNRS, Universit\'e Paris-Sud, Universit\'e Paris-Saclay, F-91406 Orsay Cedex, France}
 
\date{\today}
\begin{abstract} 
We further progress along the line of Ref. [D. Lacroix, Phys. Rev. {\bf A 94}, 043614 (2016)] where a 
functional for Fermi systems with anomalously large $s$-wave scattering length $a_s$ was proposed that has 
no free parameters. The functional is designed to correctly reproduce the unitary limit in Fermi gases together with 
the leading-order contributions in the s- and p-wave channels at low density. The functional is shown to be predictive 
up to densities $\sim0.01$ fm$^{-3}$ that is much higher densities compared to the Lee-Yang functional, valid 
for $\rho < 10^{-6}$ fm$^{-3}$. The form of the functional retained in this work is further motivated. It is shown that the new functional 
corresponds to an expansion of the energy in $(a_s k_F)$ and $(r_e k_F)$ to all orders, where $r_e$ is the effective range and 
$k_F$ is the Fermi momentum.
One conclusion from the present work is that, except in the extremely low--density regime, 
nuclear systems can be treated perturbatively in $-(a_s k_F)^{-1}$ with respect to the unitary limit.  
Starting from the functional, we introduce density--dependent scales and show that scales associated to the bare interaction 
are strongly renormalized by medium effects. As a consequence, some of the scales at play 
around saturation are dominated by the unitary gas properties and not directly to low-energy constants.
For instance, we show that the scale in the s-wave channel around saturation is proportional to the so-called Bertsch parameter $\xi_0$ and becomes independent of $a_s$. 
We also point out that these scales are of the same order of magnitude than those empirically obtained in the Skyrme energy 
density functional. We  finally propose a slight modification of the functional such that it becomes accurate up to the saturation density
$\rho\simeq 0.16$ fm$^{-3}$. 
\end{abstract}

  

\maketitle

\section{Introduction}

In the last 50 years, nuclear theoretical physics has encountered two major breakthrough. The first one was the nuclear Density Functional 
Theory (DFT) approach also called Energy Density Functional (EDF) theory. 
In the seventies, it was realized that simple functionals
 \cite{Vau72,Gog75,Gog75-b}  based on the concept of effective interaction can be very accurate while simultaneously unifying the description 
 of nuclear structure \cite{Ben03,Sto07}, nuclear dynamics \cite{Sim10,Nak16} or thermodynamics \cite{Oer16}. 
 Nuclear EDF remains, even today, the only microscopic approach 
 able to describe nuclear systems from small masses ($N\ge16$) 
 to infinite nuclear matter.
 {\it Still, the understanding of "why functional of
 extreme simplicity can work so well despite the known complexity of the underlying many-body interaction?"}
 remains unclear. 
  
 A second breakthrough was made more recently on the nuclear interaction itself and on its application to nuclear systems.  
 In particular, it was realized that the strong nuclear repulsion at short distances can be replaced by a softer potential
 that is optimized for the low-energy scales relevant for nuclei \cite{Ham09,Kuo16}. 
 Progress has been made along this line in the past decades, 
 especially under the impulse of the nuclear Effective Field Theory (EFT) leading to a constructive 
 approach \cite{Kol02,Hol13,Gez13,Gez14,Epe15,Mac16,Mei16} for the nucleon-nucleon interaction. 
 The possibility to get rid of the strong repulsion turns out to considerably simplify
 the nuclear many-body problem. In particular, nuclei become more perturbative and methods that were extremely difficult 
 to apply with former generations of bare interactions become manageable. Considerable efforts are made nowadays to 
 develop accurate exact calculations, 
 called ab-initio methods, for infinite matter and nuclei \cite{Kuo16,Hag16,Nav16,Gan15}.  A very important aspect 
 of the strategy used in ab-initio methods, is that the complexity of the nuclear interaction is gradually increased using power counting analysis 
 leading to exact calculations with controlled errors. One difficulty that is encountered is that the EFT approach 
 automatically leads to three-body and more generally many-body interactions that are not easy to handle in applications. Nevertheless, 
 by treating the Lagrangian at increasing orders, one should reduce gradually the errorbars in exact calculations. One can note however 
 that, in practice, errorbars do not decrease so fast from LO to N2LO or N3LO, etc... In addition, when applying ab-initio methods 
 to infinite systems, these errorbars are increasing with density (small relative distances), as expected, and turn out to be 
 rather large around the saturation density (see for instance \cite{Dri16}).    

 Quite naturally, attempts have been made to take advantage of these progress on the nuclear interaction and to
 obtain less empirical nuclear EDF. This includes the use of Density Matrix Expansion (DME) 
 \cite{Neg72,Geb10,Sto10,Dob10,Car10,Dyh16} 
 or functionals deduced for instance in infinite systems, like in the ongoing effort summarized in Ref. \cite{Hol13}. Another path that is now 
 explored is to clarify the notion of beyond mean-field approaches within EDF and eventually propose new functionals using techniques
from EFT \cite{Mog10,Kai15,Yan16a}.  A common feature 
 of these attempts is that the functionals become rapidly rather complicated and therefore are at variance with the apparent 
 simplicity of more empirical nuclear EDF. 
   
 Alternatively, it was noted that due to the very large $s$-wave scattering length of the nuclear interaction, nuclear systems 
are rather close to unitary gas where the scattering length becomes infinitely large. Numerous  experimental and theoretical 
works have been made on the unitary regime \cite{Gio08,Blo08,Zwe11}.  Unitary gases are universal in the sense that, independently 
from the original interaction and particles nature, the energy can be written as $E= \xi_0 E_{\rm FG}$ where $\xi_0$ is the so-called Bertsch parameter \cite{Ber00} and 
$E_{\rm FG}$ is the free gas energy.  In view of this, minimalist functionals have been produced at unitarity \cite{Pap05,Bul07,Zwe11}.  

Our primary goal is to construct simple functionals for neutron matters with as less free parameters as possible, rendering the functionals 
at the same time less empirical. The starting point in the recent work \cite{Lac16} is to use the universality of unitary gas on one side and the 
behavior of nuclear systems at very low density to propose a new functional. Here, we report new progress we have made along this line. A more precise discussion and justification of the functional form we have retained 
is made. The functional is finally further improved by imposing that the effective range dependence of the unitary gas  is better reproduced.

The novel functional leads naturally to density--dependent scales that identify with the bare scales at very low density and strongly evolve with the density.  We show that the scale obtained in this way helps to understand why empirical functionals like Skyrme 
based EDF, although very simple and not connected to the bare interaction 
can be so predictive (see Section \ref{sec:ddlec}). 

\section{Parameter free nuclear Density Functionals}

Following the recent work of ref. \cite{Lac16}, we focus here on neutron matter. We consider a spin-degenerate system interacting through 
an s-wave interaction characterized by its phase shift $\delta_0$ at low momentum transfer:
\begin{eqnarray}
k \cot \delta_0 &=& - \frac{1}{a_s} + \frac{1}{2} r_e k^2 + O(k^4).
\end{eqnarray}
where $k$ is the relative momentum of the interacting particles and where the low--energy constants (LECs) $a_s$ and $r_e$ stand respectively for the s-wave scattering length and the effective range. For neutron matter, these low--energy constants are equal to $a_s=-18.9$ fm and 
$r_e=2.7$ fm.

Guided by the resummation technique used in low-density Fermi gases with large scattering length \cite{Ste00,Sch05,Kai11} and on the recent efforts to develop a nuclear energy density functional correctly treating low--density fermi liquids \cite{Yan16}, a novel density functional for neutron matter was proposed in Ref.  \cite{Lac16}. This functional  can be generically written as:
\begin{eqnarray}
\frac{E}{E_{\rm FG}} &=& \xi(a_s k_F, r_e k_F) \label{eq:xireas}
\end{eqnarray}
where $E_{\rm FG}$ is the free Fermi gas energy given by $E_{\rm FG }/ N = 3\hbar^2k_F^2/(10 m)$.
Here $N$ is the number of particles, $k_F$ is the Fermi momentum that is obtained from the single-particle density $\rho$ through $\rho=\nu k^3_F/(6\pi^2)$. $\nu$ is the degeneracy ($\nu=2$ for neutron matter). In Ref. \cite{Lac16},  the following $\xi$ functional has been proposed:
\begin{eqnarray}
 \xi(a_s k_F, r_e k_F) = 1 + \frac{5}{3}  \frac{(a_s k_F) A_0  }{1 - A_0^{-1} \left[ A_1 + (r_e k_F)A_2   \right] a_s k_F} \label{eq:xi1}
 \end{eqnarray}
Besides this specific form, an important aspect is that the parameters $\{ A_i\}$ are not adjustable but are fixed by
imposing well-defined limits.

\begin{figure*}[htbp]
\includegraphics[width=\linewidth]{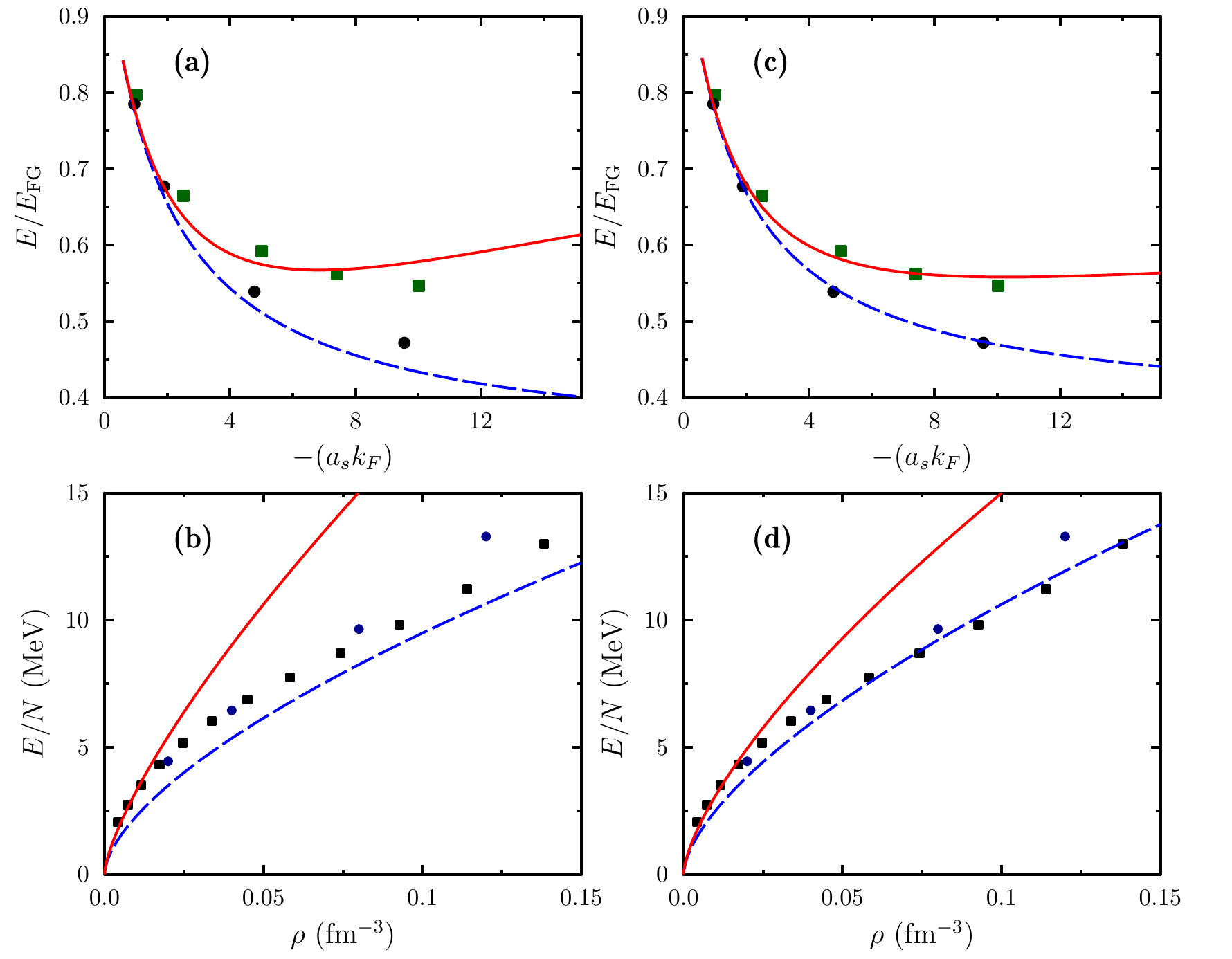}
\vspace{-5mm}
\caption{ (color online) Neutron matter energy as a function either of $-(a_s k_F)$ or $\rho$. Panels (a) and (b) correspond to 
the results obtained by using Eq. (\ref{eq:xi1}) where the ${A_i}$ coefficients are adjusted by imposing the developments given by Eqs. (\ref{eq:leeyanghf}-\ref{eq:leeyang2nd}). 
Panels (c) and (d) correspond to Eq.  (\ref{eq:funcunit}) where $U_i$ and $R_i$ are obtained using Eq. (\ref{eq:leeyanghf}) and unitarity limits as constraints. 
In all panels, the red solid lines correspond to the equation of state including the effective range dependence while the blue dashed lines correspond to the result obtained by assuming $r_e=0$. In panels (a) and (c), the black circles and green squares correspond to the QMC results of Ref. \cite{Gez10,Car12} respectively for the cold atom case and AV4 (s-wave only) case. In panels (b) and (d), the black squares and darkblue circles correspond respectively to the ab-initio results of Ref. \cite{Fri81} and Ref. \cite{Akm98}. Note finally that the highest density shown in
the upper panels corresponds to $\rho=0.01$ fm$^{-1}$ and corresponds to a very narrow region of panels (b) and (d). 
}
\label{fig:func1} 
\end{figure*}

One possibility that has been explored in Refs. \cite{Sch05,Yan16} is to fix some parameters by imposing the  correct low--density limit. In general, the energy of a Fermi system can be written as : 
\begin{eqnarray}
\frac{E}{E_{\rm FG}}&=&1 + \frac{ E^{(1)}}{E_{\rm FG}} +  \frac{ E^{(2)} }{E_{\rm FG}} + \cdots \label{eq:leeyang}
\end{eqnarray}
where $E^{(1)}$ is the Hartree-Fock energy, $E^{(2)}$ (resp. $E^{(n)}$) is the second-order (resp. $n^{th}$-order)
perturbation theory contribution.  At low density, the different contributions can be expanded in power of $k_F$ as \cite{Fet71} 
\footnote{Note that these expressions are valid if pairing correlations are neglected. The presence of pairing estimated using the Hartree-Fock Bogolyubov approach \cite{Fet71} would lead to an additional contribution associated for instance in the s-wave channel to a pairing gap $\Delta \propto \exp\left(- \pi/[2 k_F |a_s|]\right)$ \cite{Pap99}. }:
\begin{eqnarray}
\frac{ E^{(1)}}{E_{\rm FG}} &=&  \frac{10}{9\pi}(\nu -1) (k_F a_s)
+  (\nu- 1) \frac{1}{6 \pi} (k_F r_e) (k_F a_s)^2  \nonumber \\  
&+& (\nu+1) \frac{1}{3 \pi} (k_F a_p)^3 + \cdots  \label{eq:leeyanghf} \\
\nonumber \\
\frac{ E^{(2)}}{E_{\rm FG}} &=& (\nu -1) \frac{4}{21 \pi^2} (11-2 \ln2)(k_F a_s)^2 + \cdots \label{eq:leeyang2nd}
\label{eq:degnm}
\end{eqnarray}
We recognize in particular some of the terms appearing in the Lee-Yang formula obtained in Refs. \cite{Lee57, Bis73,Ham00}.
Setting the p-wave scattering volume to zero and imposing to recover the different terms appearing in Eqs. (\ref{eq:leeyanghf}-\ref{eq:leeyang2nd})
when Taylor expanding (\ref{eq:xi1}) to second order in $(a_s k_F)$ and first order in $(r_e k_F)$  provides a unique determination of the $A_i$ parameters (see \cite{Lac16} for explicit values).  Results of this method are shown in panels (a) and (b) of Fig. \ref{fig:func1}. As noted in Ref. \cite{Sch05,Lac16}, these results are actually not so far from the "exact" QMC at low density even if $-(a_sk_F)\gg 1$. 

As an alternative strategy, starting from the fact that the s-wave scattering length in nuclear matter is very large, it was proposed to constrain the functional (\ref{eq:xi1}) keeping the constraint of the Hartree-Fock expansion, Eq. (\ref{eq:leeyanghf}) while
using the unitary gas limit 
instead of the second--order contribution. The unitary regime corresponds to the limit $-(a_sk_F)^{-1} \rightarrow 0$.  After simple manipulations, the unitary gas limit is better emphasized by rewriting Eq. (\ref{eq:xi1}) as:
 \begin{eqnarray}
&&\frac{E}{E_{\rm FG}}= 1 -  \frac{U_0  }{1 -  (a_s k_F)^{-1} U_1 }  \nonumber \\  
&& \hspace*{0.cm} +  \frac{R_0  (r_e k_F) } {\left[ 1- R_1  (a_s k_F)^{-1} \right] 
\left[ 1-R_1  (a_s k_F)^{-1} + R_2  (r_e k_F)\right]}  \label{eq:funcunit}
\end{eqnarray}
where new parameters can be expressed in terms of the original $A_i$ coefficients\footnote{It can be shown that starting from Eq. (\ref{eq:xi1}), we have the relationships
$U_1=R_1$ and $R_0  =  U_0 R_2$, so that the number of independent parameters in (\ref{eq:funcunit}) is the same as in (\ref{eq:xi1}).  However, in the present study, we will not impose this constraint and simply fix the 5 parameters entering in Eq. (\ref{eq:funcunit}) independently from each other.}. 
In Ref. \cite{Lac16}, the three independent parameters have been adjusted by imposing that the leading order in the low--density 
expansion in $(a_s k_F)$ and the behavior of the quantity $E/E_{\rm FG}$ at unitarity are correctly reproduced. More precisely,  
we took advantage of the recent study \cite{For12} where the possible effect of non-zero effective range 
in unitary gases was analyzed:
\begin{eqnarray}
\xi(+\infty, r_ek_F) &\equiv &\xi_0 + ( r_e k_F) \eta_e   + ( r_e k_F)^2\delta_e . \label{eq:etae}
\end{eqnarray}
$\xi_0$ is the Bertsch parameter while $\eta_e$ and $\delta_e$ are two new parameters. In the present work, 
we take the reference values $\xi_0=0.376$, $\eta_e=0.127$, and $\delta_e= - 0.055$ \cite{For12}. It is worth mentioning
that these values correspond to averages over the different interactions considered in Ref. \cite{For12}. We tested the sensitivity 
of the result to the $\eta_e$ value (assuming $\delta_e=0$). Reducing $\eta_e$ to 0.046 as originally obtained in Ref. \cite{Pap06} gives a 
slightly lower energy at low density while the shape and order of magnitude of the energy is globally unchanged for the density considered in panel (d) of Fig. \ref{fig:func1}.

In our previous work, accounting for the constraints between the $U_i$ and $R_i$, three constraints were necessary to fix
the 3 independents parameters. Then,  only $\xi_0$ and $\eta_e$ where used as a constraint \cite{Lac16} together with the correct leading order (LO) 
in $(a_s k_F)$ at low density. 
Here we slightly improve the functional by directly using 
expression (\ref{eq:funcunit}) and relax the constraints between the different $U_i$ and $R_i$ coefficients. 
Then 5 constraints are needed to fix the 5 
parameters. 
We impose that the three terms of Eq. (\ref{eq:etae}) are reproduced as well as the the second and third terms of Eq. (\ref{eq:leeyanghf}).
This gives: 
\begin{eqnarray}
\left\{
\begin{array}{l}
\displaystyle U_0 =  (1-\xi_0) = 0.62400,  \\
\displaystyle U_1 = \frac{9 \pi}{10} (1-\xi_0)  =  1.76432, \\
\displaystyle R_0 = \eta_e= 0.12700,   \\
\displaystyle R_1 = \sqrt{\frac{6 \pi \eta_e}{(\nu -1)}}= 1.54722, \\
\displaystyle R_2 = - \delta_e/\eta_e= 0.43307. 
\end{array}
\right.
\label{eq:param}
\end{eqnarray}
Results of the functional (\ref{eq:funcunit}) are shown in panels (c) and (d) of Fig. \ref{fig:func1}.  

In panel (c) of Fig. \ref{fig:func1}, we also display the result obtained with the functional  (\ref{eq:funcunit}) assuming $r_e=0$ (blue dashed line) and compare it with the QMC calculations obtained in Ref. \cite{Gez10,Car12} (filled circles). In this case, the functional  only depends on the two parameters $U_0$ and $U_1$ that are both only functions of the Bertsch parameter $\xi_0$. Despite its simplicity the functional result perfectly matches the exact QMC result.  As noted in Ref. \cite{Lac16}, it confirms the finding of Refs. \cite{Adh08,Adh08-b} where a similar functional was proposed at unitarity.
 
By comparing the two red lines displayed respectively in panels (a) and (c) of Fig. \ref{fig:func1} with the QMC results in neutron matter (filled squares), we also note that taking the constraint on the unitary limit instead of the constraint to reproduce the term appearing in (\ref{eq:leeyang2nd}) significantly improves the description of neutron matter. From this, it seems quite clear that the unitary gas regime is a good starting point.

\begin{figure}[htbp]
\includegraphics[width=\linewidth]{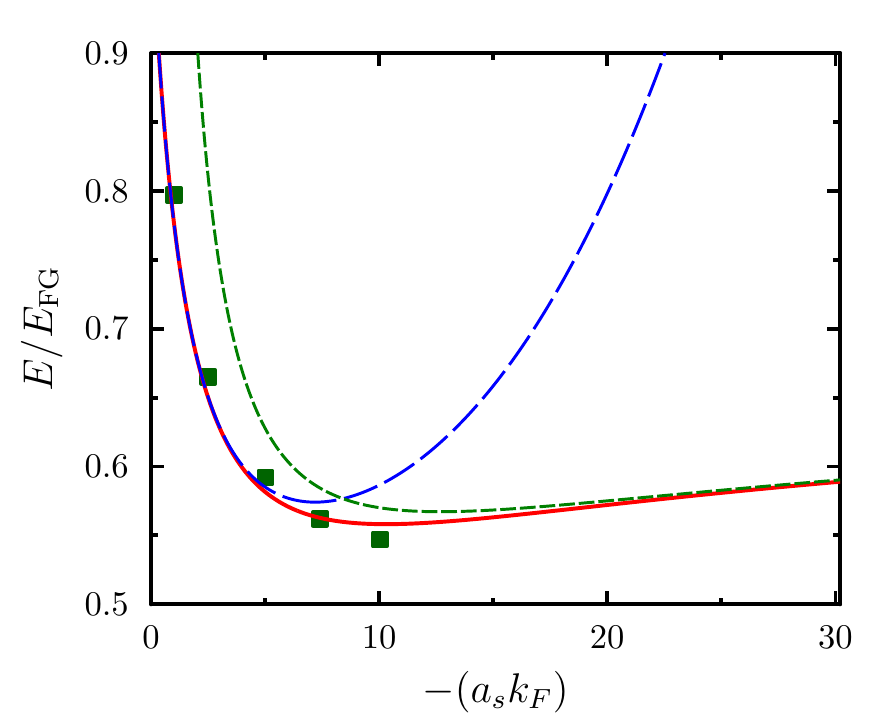}
\vspace{-5mm}
\caption{ (color online)  Same as panel (c) of Fig. \ref{fig:func1}. The green squares correspond to QMC results of Ref. \cite{Gez10} and the red solid line to results of Eq.  (\ref{eq:funcunit}). 
The green short dashed line and blue long dashed line correspond to the expansion of Eq. (\ref{eq:funcunit}) 
to first--order in $(a_s k_F)^{-1}$ or to first--order in $(r_e k_F)$ respectively. }
\label{fig:dev} 
\end{figure}

In order to illustrate how perturbative 
is neutron matter with respect to unitary gas and/or low--density regime, 
we show in Fig. \ref{fig:dev} the results of first--order Taylor 
expansion of Eq. (\ref{eq:funcunit}) either in $(a_s k_F)^{-1}$ or in $(r_e k_F)$. 
It is then clear that the former expansion rapidly converges to the full expression. 
In particular, at densities above $\rho=0.01$ fm$^{-3}$, the first--order expansion in $(a_s k_F)^{-1}$ cannot be distinguished 
from the result of Eq. (\ref{eq:funcunit}). 
It is interesting also to mention that the first--order expansion in $(r_e k_F)$ deviates rather fast from the 
function (\ref{eq:funcunit}). Indeed, although $r_e$ is much smaller than $|a_s|$, for densities around 
the saturation density of symmetric matter, we are beyond the range of validity of an expansion in the effective range.
It can also be stressed that Eq. (\ref{eq:funcunit}) contains all orders in  $(r_e k_F)$ and therefore can also be seen as a resummed 
expression accounting for effective range effects. This aspect is discussed below. 

\section{Mean-field based on finite-range interaction: discussion of resummation of effective range effect}

In Ref. \cite{Lac16}, the effective range dependence of the unitary gas regime has been introduced 
without much justifying the retained expression. Here, we would like to give more physical insight 
in the expression and to show that the effective range dependence entering in (\ref{eq:xi1}) can be also  regarded 
as a resummation of $r_e$ effect.  The easiest way to introduce effective range effects beyond those contained in Eq. 
(\ref{eq:leeyanghf}) is to consider the Hartree-Fock energy  associated to a finite-range interaction. We take below the case of a Gaussian interaction. This section also illustrates how the interaction parameters should be adjusted to properly account for the low--energy 
constants coming from the underlying Lagrangian. This strategy is very similar to what is usually done in Effective Field Theory (EFT).  The connection with EFT based on zero range interaction is then naturally made here.  Finally, a discussion on 
the possibility to make resummation of the effective range such that the functional 
(\ref{eq:xi1}) is recovered. 

\subsection{Preliminary: Hartree-Fock energy of a Gaussian two-body interaction}

We consider here  a two-body gaussian interaction written in the form:
\begin{eqnarray}
v({\mathbf r_1},{\mathbf r_2}) &=&  \Big\{ v_{0}+ v_\sigma
P_\sigma \Big\}  g({\mathbf r}_1 - {\mathbf r}_2)  \label{eq:gaussian}
\end{eqnarray}
where $v_0$ and $v_\sigma$ are parameters. 
$P_\sigma$ is the operator that exchanges the spin of two particles and $g$ is a normalized Gaussian given by:
 \begin{eqnarray}
g( r ) = \frac{e^{-(r^2/\mu^2)}}{(\mu \sqrt{\pi})^3},
\end{eqnarray}
where $r=|{\mathbf r_1}-{\mathbf r_2}|$ and $\mu$ is a free parameter.
The Hartree-Fock contribution of 
this interaction to the equation of state of neutron matter gives:
\begin{eqnarray}
 \frac{E^{(1)}_G}{N} & = & \frac{\rho}{2} \left[ A  - B F\left( \mu k_F \right) \right], \label{eq:egaus}
 \end{eqnarray}
 where $A$ and $B$ are
 \begin{eqnarray}
A & = & v_0 + \frac{1}{2} v_\sigma, ~~~~~~
B = \frac{1}{2} v_0 + v_\sigma. \label{eq:ab}
\end{eqnarray} 
$F$ is a function given by: 
\begin{eqnarray}
F(x) =  \frac{4}{\sqrt{\pi} x^3} \int_0^{+\infty}  z^2 \mathrm{e}^{-z^2/x^2} \left(\frac{3j_1(z)}{z} \right)^2 \mathrm{d}{z}, \label{eq:f1} \\
=  \frac{12}{x^6} (1  -  \mathrm{e}^{- x^2}) + \frac{6}{x^4}(\mathrm{e}^{- x^2}-3) +  \frac{6\sqrt{\pi}}{x^3} \mathrm{Erf}{(x)} , \label{eq:f2}
\end{eqnarray}
where $j_1(z)$ denotes the spherical Bessel function and $x= \mu k_F$. 
 
\subsection{Discussion on the low--density limit and relation between the EFT, Skyrme and Gaussian interaction parameters}
\label{sec:lec} 

Fermi systems at low density have been widely studied using the EFT approach
based on a zero-range interaction. In that case, spin--degenerate systems can be studied using a two-body
interaction written in the form \cite{Ham00}:
\begin{eqnarray}
\langle {\mathbf k} |V_{\rm EFT} |  {\mathbf k'} \rangle = C_0 + \frac{C_2}{2} \left(  {\mathbf k}^2  +  {\mathbf k'}^2 \right) + C'_2 ~{\mathbf k}.{\mathbf k}' ,
\end{eqnarray} 
or equivalently in $r$-space:
\begin{eqnarray}
v({\mathbf r}_1,{\mathbf r}_2) &=& C_0 \delta({\mathbf r}_1 -{\mathbf r}_2) \nonumber \\
&+& \frac{1}{2} C_2 \left[  \delta({\mathbf r}_1 -{\mathbf r}_2){\mathbf k}^2 + {\mathbf k'}^2\delta({\mathbf r}_1 -{\mathbf r}_2) \right] \nonumber \\
&+& C'_2 {\mathbf k}'  \delta({\mathbf r}_1 -{\mathbf r}_2) {\mathbf k}' . \label{eq:eft0}
\end{eqnarray} 
Using this interaction, the Hartree-Fock energy then read:
\begin{eqnarray}
\frac{E}{N} & = & \frac{3}{5} \frac{\hbar^2 k^2_F}{2m} \nonumber \\
&+& \frac{k^3_F}{4 \pi^2} \left( \frac{C_0}{3} + \frac{k_F^2}{10} \left[(\nu-1)C_2  + (\nu+1)  C'_2  \right] \right) .\label{eq:eft1}
\end{eqnarray} 
At low--density, we can then compare to the Hartree-Fock contribution given in Eq. (\ref{eq:leeyanghf}). 
The low density expansion of the HF energy is recovered under the condition that the parameters $C_i$ are linked to the low--energy constants 
through \cite{Ham00}:
\begin{eqnarray}
 C_0 = \frac{ 4 \pi \hbar^2}{m}  a_s, ~~
 C_2 = \frac{2 \pi \hbar^2 }{m} r_e  a_s^2,~~
 C'_2 =\frac{4 \pi \hbar^2}{m}  a_p^3. 
\end{eqnarray} 
$a_p^3$ is the so-called p-wave scattering volume \footnote{Note that here we use a slightly different notation compared to standard definition. Indeed, the p-wave scattering volume $\alpha_p$ is usually defined though the $l=1$ phase-shift using:
\begin{eqnarray}
\lim_{k \rightarrow 0} k^3 \cot(\delta_1) = -\frac{1}{\alpha_p}
\end{eqnarray}
We use here $\alpha_p = a^3_p$ so that $a_p$ has a length unit.
 }.
Similarly to the EFT case based on zero-range interaction, one can recover the low density expression starting from the Gaussian interaction. 
The most direct way to make connections between Eq. (\ref{eq:gaussian}) and Eqs. (\ref{eq:eft0}) $\&$ (\ref{eq:eft1}) is to expand the interaction in terms of the 
momentum transferred. Starting from Eq. (\ref{eq:gaussian}) and expanding in $({\mathbf k} - {\mathbf k}')$ up to second order, we have:
\begin{eqnarray}
v ({\mathbf k}, {\mathbf k}') &=& 
\left( v_{0}  + v_\sigma  P_\sigma  \right) \left( 1 - \frac{\mu^2}{4} \left[ {\mathbf k}^2  + {{\mathbf k}'}^2 \right] \right) \nonumber \\
&+& \frac{1}{2}  \mu^2  \left( v_{0} + v_\sigma  P_\sigma \right)  {\mathbf k}.{\mathbf k}' + \cdots  \label{eq:expand2}
\end{eqnarray}  
Making the inverse Fourier transform, one then obtains the Skyrme--type interaction with standard $(x_i,t_i)$ parameters:
\begin{eqnarray}
v({\mathbf r}_1 ,{\mathbf r}_2) &=& t_0 (1+x_0 P_\sigma) \delta({\mathbf r}_1 -{\mathbf r}_2) \nonumber \\
&+& \frac{1}{2} t_1 (1 + x_1 P_\sigma) \left[  \delta({\mathbf r}_1 -{\mathbf r}_2){\mathbf k}^2 + {\mathbf k'}^2\delta({\mathbf r}_1 -{\mathbf r}_2) \right] \nonumber \\
&+& t_2  (1 + x_2 P_\sigma)  {\mathbf k}' . \delta({\mathbf r}_1 -{\mathbf r}_2) {\mathbf k}' + \cdots 
\end{eqnarray}
where we have set 
\begin{eqnarray}
t_0 & = & v_0 , ~~  t_0 x_0  =  v_\sigma, \\
t_1 &=&-\frac{\mu^2}{2} v_0 , ~~
t_1 x_1 = -\frac{\mu^2}{2} v_\sigma ,  \\
t_2 &=& \frac{\mu^2}{2} v_{0}, ~~t_2 x_2 = \frac{\mu^2}{2} v_\sigma .
\end{eqnarray}
These relationships on the one side between the Skyrme parameters and the parameters of the Gaussian 
and, on the other side, the evident similarities between Skyrme and EFT Hamiltonians is a useful 
guidance to understand how the low--density behavior can be correctly reproduced using the Gaussian interaction. 
For instance, starting from Eq. (\ref{eq:egaus}), the standard expression of the neutron matter energy is recovered 
by expanding the function $F$ up to second order in $(\mu k_F)$. This expression matches Eq. (\ref{eq:eft1}) and Eq. (\ref{eq:leeyanghf})  under the set of conditions:
\begin{eqnarray}
C_0 &=& t_0 (1-x_0) = v_0 - v_\sigma =\frac{ 4 \pi \hbar^2}{m}  a_s ,   \label{eq:lec1} \\
C_2 &=&t_1 (1-x_1) = -\frac{\mu^2}{2} (v_0 - v_\sigma)  = \frac{2 \pi \hbar^2 }{m} r_e  a_s^2 , \label{eq:lec2} \\
 C'_2 &=&t_2(1+x_2) = \frac{\mu^2}{2} (v_0 + v_\sigma) =\frac{4 \pi \hbar^2}{m}  a_p^3 . \label{eq:lec3} 
\end{eqnarray}
With these relations, the same energy is obtained at the Hartree-Fock level either using the EFT Lagrangian, 
the Skyrme Hamiltonian or the Gaussian interaction provided that the energy  (\ref{eq:egaus}) is expanded up to 
second order in  $(\mu k_F)$.

There are several useful relations that could be derived for the Gaussian interaction when the low--density 
limit is imposed. For instance, we have:   
\begin{eqnarray}
\mu^2 & = & -2 \frac{C_2}{C_0}= - (r_e a_s) . \nonumber
\end{eqnarray}   
We then also get that the two parameters $A$ and $B$ entering in Eq. (\ref{eq:egaus}) respectively write:
\begin{eqnarray}
A & = & -\frac{2 \hbar^2 \pi}{\nu m \mu^2} \left[ (\nu-1) r_e a^2_s - 2 (\nu+1) a_p^3\right],  \nonumber \\
B & = &  +\frac{2 \hbar^2 \pi}{\nu m \mu^2} \left[ (\nu-1) r_e a^2_s + 2 (\nu+1) a_p^3\right].  \nonumber
\end{eqnarray}
  
\subsection{Resummation of  effective range effect in Hartree-Fock theory}
\label{sec:gauss}

One motivation of the introduction of a finite-range interaction instead of a zero-range ansatz is 
the possibility to explore the effect of higher powers of the effective range, at least in the Hartree-Fock energy.
For instance, Eq. (\ref{eq:egaus}) contains all powers in $(\mu k_F)$. Our objective here is to show that the 
approximation (\ref{eq:xi1}) can be inferred from the Gaussian interaction case.

Starting from  (\ref{eq:egaus}) and using the Taylor expansion of $F$, we deduce:
\begin{eqnarray}
 \frac{E^{(1)}_G}{N} & \simeq & \frac{\rho}{2} \left[ A  - B \left\{ 1 - \frac{3}{10} (\mu k_F)^2 \right\} \right] , \nonumber \\
 &\simeq&  \frac{\rho}{2} 
 \left[ A  - \frac{B}{1 + \frac{3}{10} (\mu k_F)^2}  \right]  %
 \end{eqnarray}
For simplicity, we now set the p-wave scattering volume to zero. In this case, we have:
\begin{eqnarray}
\frac{\rho}{2} A  & = & - \frac{\rho}{2} B =  \left( \frac{E_{\rm FG}}{N}\right)  \frac{5}{9 \pi}  (\nu-1)  (k_F a_s).
\end{eqnarray}
Then, using the value of $\mu$, we deduce:
\begin{eqnarray}
 \frac{E^{(1)}_G}{E_{\rm FG}}  =  
 \left( \frac{5}{9 \pi} \right) [\nu-1] \left( a_s k_F\right) \left\{ 1 +\frac{1}{1 - \frac{3}{10} (a_s k_F) (r_e k_F)} \right\}. \label{eq:resum0}
\end{eqnarray}
  
\begin{figure}[htbp]
\includegraphics[width=\linewidth]{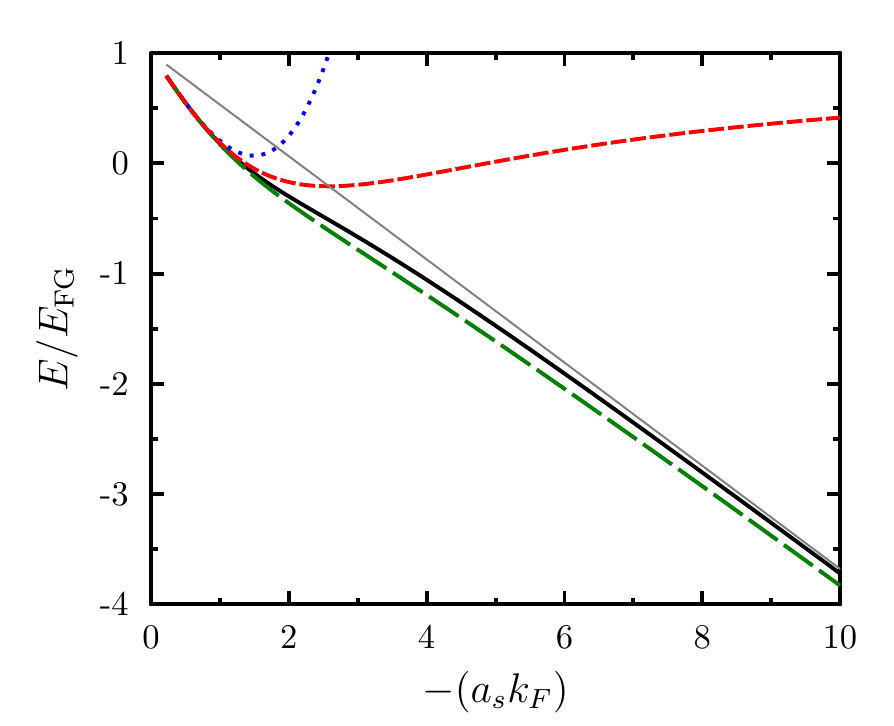}
\vspace{-5mm}
\caption{ (color online) $E/E_{\rm FG}$ as a function of $-(a k_F)$ obtained for a Gaussian interaction. The thick black solid line corresponds to Eq. (\ref{eq:egaus}) where $A$ and $B$ are calculated from low--energy constant (with $a^3_p=0$). Note that the energy 
$E$ corresponds to the sum of the kinetic and Hartree-Fock energy. The blue dotted line stands for the expansion (\ref{eq:leeyanghf}) result, while the green 
long dashed line is obtained by using Eq. (\ref{eq:resum0}).  The red short dashed line is obtained by using the alternative 
Eq. (\ref{eq:resum1}). The thin solid gray line is the function $1+5/(9\pi) (a_s k_F)$ that is shown for reference.}
\label{fig:gaus} 
\end{figure}

In Fig. \ref{fig:gaus}, the equivalent of the $\xi$ parameter, Eq. (\ref{eq:xireas}), obtained for the Hartree-Fock energy 
of a Gaussian interaction is shown. In this figure, it is compared to the 
expansion (\ref{eq:leeyanghf}) and to the result of  (\ref{eq:resum0}).
Obviously, Eq. (\ref{eq:leeyanghf}) can only grasp the low--density regime of the HF energy. When using Eq. (\ref{eq:resum0}), the low--density limit 
is well reproduced. This finding has strongly guided the 
form of the functional proposed in Ref. \cite{Lac16}. Since the approximate form essentially mimics the effect of higher powers in $(r_e k_F)$,
it can be interpreted as a resummed formula of the effective range for the Hartree-Fock energy.  In particular, it is expected to have the correct limit at high density. We see however that it converges slower than the exact case to the energy $ \left( \frac{5}{9 \pi} \right) [\nu-1] \left( a_s k_F\right)$. The faster convergence in the exact Hartree-Fock stems from the gaussian that appears in Eq. (\ref{eq:f2}). 
Alternative resummed expressions of effective-range effects are underway \cite{Bou17}.  
 
 Our ultimate goal was however not to reproduce the Hartree-Fock energy of a finite-range interaction but to obtain 
 a functional that has the proper low--density limit and a finite value at unitarity. This is not the case for the functional 
 (\ref{eq:resum0}). However, resummation can be slightly modified to give:
\begin{eqnarray}
 \frac{E^{(1)}_G}{N} &\simeq&  \frac{\rho_0}{2} \frac{\left[ A  - B \right]}{1 -\frac{3 B}{10\left[ A  - B \right]} (\mu k_F)^2} . \label{eq:resum1}
 \end{eqnarray}
 If we impose to reproduce the low--density limit, we get: 
\begin{eqnarray}
\frac{\rho}{2} [ A - B] & = & \left( \frac{E_{\rm FG}}{N}\right)  \frac{10}{9 \pi}  (\nu-1)  (k_F a_s),
\end{eqnarray}
together with:
\begin{eqnarray}
\frac{B (\mu k_F)^2}{\left[ A  - B \right]}=  \left[ \frac{1}{2}(a_s k_F) (r_e k_F) + \left( \frac{\nu+1}{\nu-1} \right) \frac{(a_p k_F)^3}{(a_s k_F)} \right]. \label{eq:mixreap}
\end{eqnarray}
To make contact with Eq. (\ref{eq:xi1}), we set to zero the p-wave scattering volume and then obtain
\begin{eqnarray}
 \frac{E^{(1)}_G}{E_{\rm FG}}  & = & 
 \left( \frac{10}{9 \pi} \right) \frac{[\nu-1] \left( a_s k_F\right)}{1 - \frac{3}{20} (a_s k_F) (r_e k_F)}. \label{eq:resumx}
\end{eqnarray} 
We recognize the terms $5/3 A_0$ and $A_0^{-1}/A_2$ obtained in Eq. (\ref{eq:xi1}) when imposing the low--density limit.
Not surprisingly, the term $A_1$ is not present since it was introduced to resum effects beyond Hartree-Fock \cite{Yan16}. The result of Eq. (\ref{eq:resum1}) is shown by short dashed line in Fig. \ref{eq:xi1}. By construction, it now goes to a finite value at large $-(a_s k_F)$  
while the low--density behavior is preserved. Note that the fact we do not reproduce the exact Hartree-Fock result (for the Gaussian interaction) is not an issue since our 
ultimate goal is to treat the energy of a highly correlated system at large scattering length.  

The Gaussian example gives some phenomenological insight on how the functional was originally 
guessed. The term $A_1$ contains in some effective way many-body effects beyond Hartree-Fock while the term $A_2$ 
contains effective range effects,  both terms being interpreted as re-summation of complex many-body diagrams to all orders. 
From the present discussion, it is also clear that the specific formula used to include effective range effect is not unique. Below, we will 
show that an alternative formula can improve the density functional at higher densities.

\subsection{Inclusion of p-wave scattering volume effect in the functional}

Our aim is now to improve the functional (\ref{eq:funcunit}) by including possible p-wave effects. Let us first estimate the p-wave
scattering volume relevant for neutron matter.  Two neutrons can interact through $^{3}P_0$, $^{3}P_1$ and $^{3}P_2$.
p-wave scattering volume estimates for each of these channels can be found in Ref. \cite{Mac01,Wir95, Val04}. We retain here the values for neutron-neutron
scattering:\begin{eqnarray}
a^3_0 &\simeq & -2.45 ~{\rm fm}^3 ,\nonumber \\
a^3_1 &\simeq & 1.50 ~{\rm fm}^3 ,\nonumber \\
a^3_2 &\simeq & -0.29 ~{\rm fm}^3.
\end{eqnarray}
Here $a^3_i$ denote the p-wave scattering volume in the channel $^{3}P_i$.
Accounting for the fact 
that these channels have respectively $1$, $3$ and $5$ spin projection, the average p-wave scattering volume in neutron
matter can be estimated through the weighted average:
  \begin{eqnarray}
a^3_p= \frac{1}{9}\left(a^3_{0} + 3 a^3_{1}+ 5 a^3_{2}\right)
\end{eqnarray} 
leading to $a^3_p \simeq  0.6 ~{\rm fm}^3$. In particular, we see that we have the hierarchy of scales $a_s \gg r_e > a_p$. 
Note that the value of the scattering volume given above is directly extracted from experimental observation. It however 
differs from the p-wave scattering volume directly deduced for the AV4 interaction that was used in the QMC approach. 
For this simplified nuclear interaction, the three channels $^{3}P_0$, $^{3}P_1$ and $^{3}P_2$ are degenerate and have 
a scattering volume equal to  $0.25 ~{\rm fm}^3$. Although $0.6 ~{\rm fm}^3$ seems more appropriate for nuclear systems, when comparing 
to QMC, the value $0.25 ~{\rm fm}^3$ is more meaningful.

\begin{figure}[htbp]
\includegraphics[width=\linewidth]{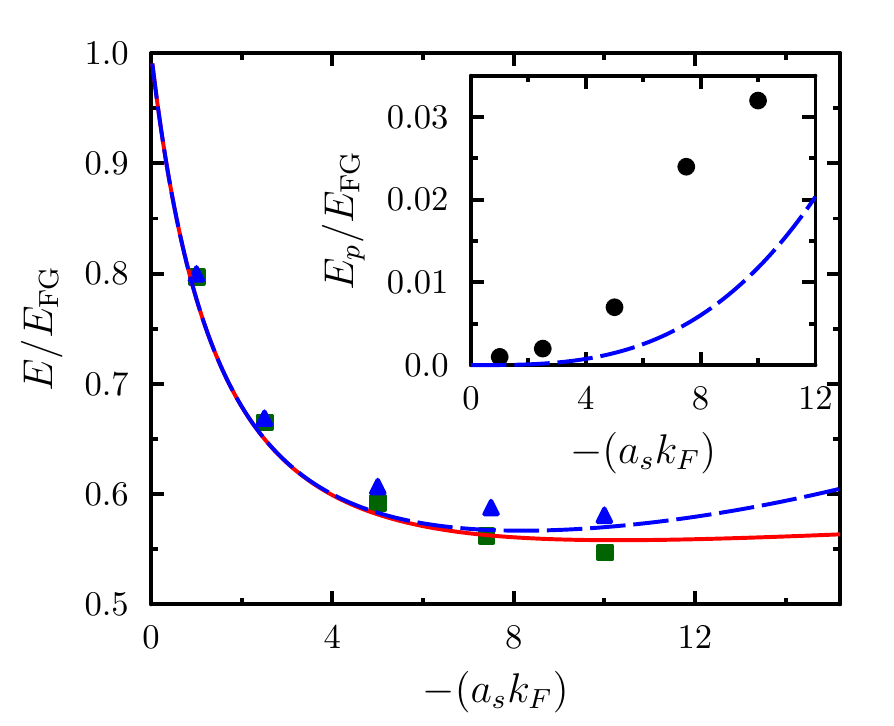}
\vspace{-5mm}
\caption{ (color online) Energy of neutron matter obtained in QMC calculations retaining only the s-wave contribution 
(green squares) or using the full AV4 interaction (blue triangle). The red solid corresponds to the result of the functional (\ref{eq:funcunit}) without the p-wave contribution. The  blue long--dashed line is obtained by adding 
the p-wave contribution using $a^3_p=0.25$ fm$^{3}$. 
In the inset, the blue long--dashed line represents the p-wave term of Eq. (\ref{eq:leeyanghf}), 
i.e. $E_p/E_{\rm FG} =  (\nu+1) (a_p k_F )^3/(3 \pi)$ 
using  $a^3_p=0.25$ fm$^{3}$. This term is compared to the difference between the energy obtained in QMC with AV4 and the QMC result with s-wave only (black filled circles).
 }
\label{fig:p-wave} 
\end{figure}

For the range of $(a_s k_F)$ considered in panel (c) of Fig. \ref{fig:func1}, we have $(a_p k_F)$ much smaller than one. 
Therefore, we anticipate that the 
third term of Eq. (\ref{eq:leeyanghf}) can eventually account for the p-wave contribution for this density range. In Fig. \ref{fig:p-wave}, we 
compare the QMC result obtained with full AV4 interaction in neutron matter \cite{Gez10} to the result obtained by 
simply adding to the functional (\ref{eq:funcunit}) the p-wave term of Eq. (\ref{eq:leeyanghf}) using $a^3_p=0.25$ fm$^{3}$.  
Similarly to the QMC calculation, we observe a global increase of the energy per particle. However, we see that the p-wave term leads to a slightly lower energy compared to QMC. This is illustrated in the inset of Fig.  \ref{fig:p-wave}. The observed difference might be due to the necessity to 
add higher multipole contributions or to the go beyond the leading order contribution for the p-wave in particular by treating 
interference terms with the s-wave channels that appear for instance due to beyond mean-field effects.

\section{Discussion of EDF for nuclear system from low to saturation density}
\label{sec:ddlec}

Following Ref. \cite{Lac16}, we are here proposing an EDF for neutron matter where parameters are determined 
either from low--energy constant of the interaction or from the unitary limit. The simplest example of such 
functional is the Lee-Yang formula with increasing numbers of terms \cite{Ham00}. Unfortunately, due to the 
large $s$-wave scattering length, such approach is restricted to very limited range of density, $\rho \le 10^{-6}$ fm$^{-3}$
that is several order of magnitude smaller than the saturation density in nuclei. We introduce here a functional (Eq. (\ref{eq:funcunit})) that seems 
to be appropriate at much higher density $\rho \le$ 0.01  fm$^{-3}$. This is still rather small 
but we have gained several orders of magnitudes. Although the functional does not treat the nuclear many-body problem 
in its full complexity and is at this stage restricted to small densities, we will show that it can bring interesting insight 
on standard functionals used currently in nuclear physics.  

One stricking aspect in EDF is the apparent simplicity and remarkable predictive power of Skyrme based EDF. The essence of Skyrme 
functionals is the use of a contact interaction where the $t_0$, $t_1$, and $t_2$ terms can be regarded as the s-wave, effective 
range and p-wave terms usually introduced in EFT (see discussion in section \ref{sec:lec}). However, as underlined in ref. \cite{Fur12},
starting from Skyrme parameters, one can estimate the equivalent values of "Skyrme LEC", that we will denote below $a^{\rm Sk}_s$, $r^{\rm Sk}_e$ and $(a^{\rm Sk}_p)^3$ using Eqs. (\ref{eq:lec1}-\ref{eq:lec3}). Deduced values have often nothing to do with physical values of nuclear
LEC (see-below). For instance, one typically obtain $|a_s| \equiv 1-5$ fm, that is much smaller than he expected 18.9 fm value. 
Here, we would like to show that the functional we propose might be useful (i) to understand why simple functional like Skyrme EDF
works so well (ii) why the equivalent LEC differs so much from the physical ones. 

Omitting density--dependent and spin-orbit term, the Skyrme EDF mean-field energy can be written as the LO energy in EFT that is given 
by Eq. (\ref{eq:eft1}), provided that we use Eqs.  (\ref{eq:lec1}-\ref{eq:lec3}). Starting from our new functional and to make contact with Skyrme or EFT, we introduce the three density--dependent terms $\widetilde C_0 (k_F)$, 
$\widetilde C_2 (k_F)$ and $\widetilde C'_2(k_F)$ and rewrite our functional as: 
\begin{eqnarray}
\frac{E}{E_{\rm FG}} & = & 1 + \frac{k^3_F}{4 \pi^2 E_{\rm FG}} \Big\{ \frac{\widetilde C_0 (k_F) }{3}  \nonumber \\
&+& \frac{k_F^2}{10} \left[(\nu-1) \widetilde C_2(k_F)  + (\nu+1)  \widetilde C'_2(k_F)  \right] \Big\}. \label{eq:dde}
\end{eqnarray} 
$\widetilde C_2(k_F)$ and $\widetilde C'_2(k_F)$ contains the term proportional to the effective range and p-wave scattering volume respectively while $\widetilde C_0(k_F)$ contains the rest. We then introduce density--dependent parameters 
$\widetilde a_s (k_F)$, $\widetilde r_e(k_F)$ and $\widetilde a^3_p(k_F)$ that are linked to the parameters $\widetilde C_{0,2} (k_F)$ and 
 $\widetilde C'_{2} (k_F)$
through relations equivalent to (\ref{eq:lec1})-(\ref{eq:lec3}). Then, the energy identifies with the expression (\ref{eq:leeyanghf}), where the LEC are replaced by the new density--dependent parameters. 

With these definitions, the density--dependent parameters can be expressed as a function of the parameters of the functional 
as:
\begin{eqnarray}
\widetilde a_s(k_F)  & = & - \frac{1} {k_F}  \frac{U_1 }{\left[1 -  (a_s k_F)^{-1} U_1 \right]}, \label{eq:astilde}
\end{eqnarray}
and 
\begin{eqnarray}
 &&  \widetilde r_e (k_F) = \frac{1}{k^3_F \widetilde a^2_s(k_F)} \nonumber \\
 && \times \frac{R^2_1  (r_e k_F) } {\left[ 1- R_1  (a_s k_F)^{-1} \right] 
\left[ 1-R_1  (a_s k_F)^{-1} + R_2  (r_e k_F)\right]}.  \label{eq:retilde} 
\end{eqnarray}
Here the $U_i's$ and $R_i's$ are listed in Eq. (\ref{eq:param}). Note finally that we simply have here $\widetilde a_p = a_p$.

\begin{figure}[htbp]
\includegraphics[width=0.8\linewidth]{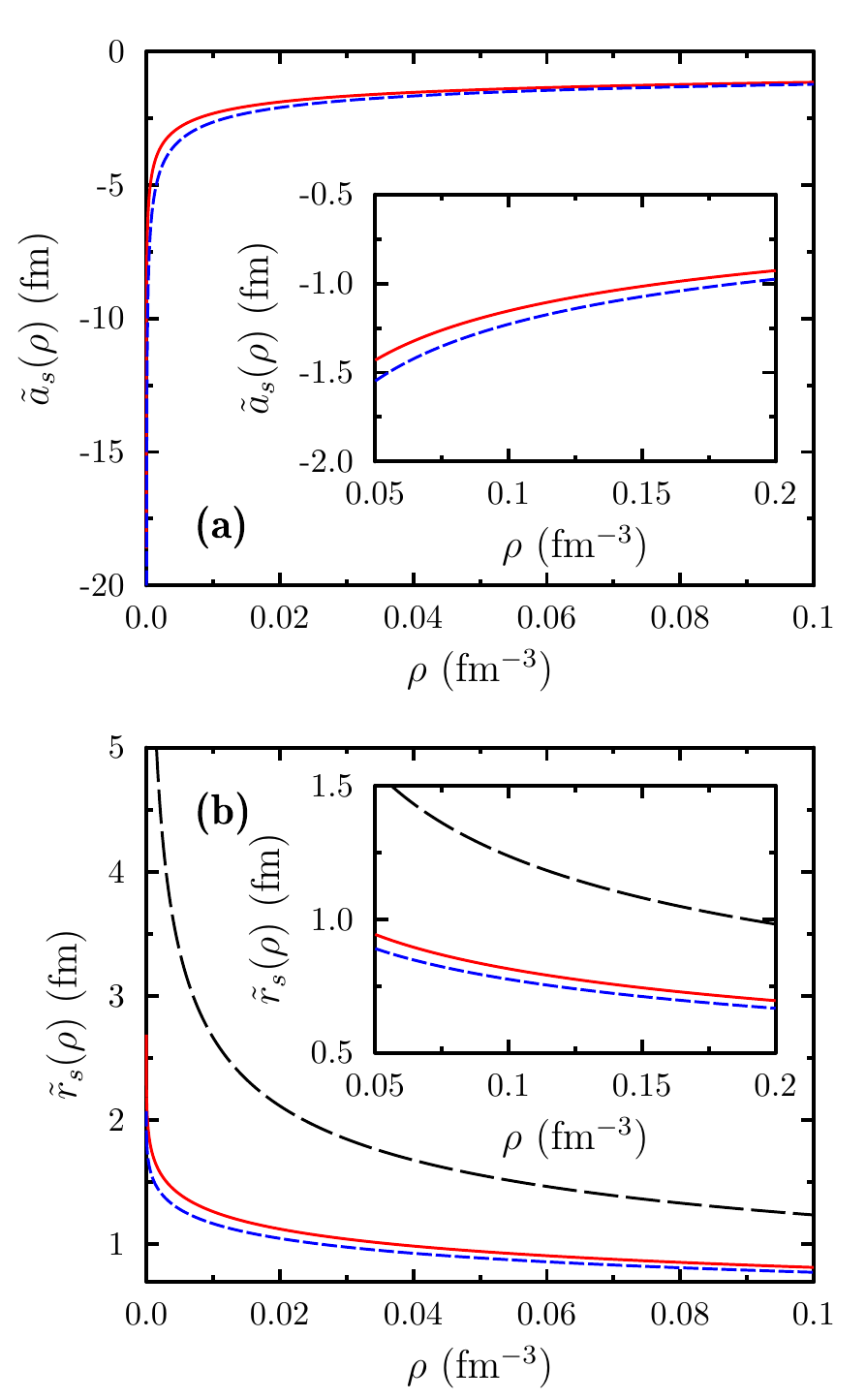}
\vspace{-5mm}
\caption{The red solid lines give the evolution of the quantities (a) $\tilde a_s(\rho)$  and (b) 
$\tilde r_e(\rho)$ calculated through the expressions (\ref{eq:astilde}) and (\ref{eq:retilde}) 
respectively. The blue short dashed lines represent the approximate expressions (\ref{eq:asinf}) [panel (a)] 
and (\ref{eq:reinf}) [panel (b)]. In panel (b), the asymptotic limit at high density given by Eq. (\ref{eq:reinf2}) 
is displayed by black long-dashed line. In both panels, the insets focus on the density region $0.05$ fm$^{-3}\le  \rho \le0.2 $ fm$^{-3}$.}
\label{fig:asretilde} 
\end{figure}

By construction, the constants $\widetilde a_s(k_F)$ and $\widetilde r_e (k_F)$ tend to the physical LEC 
at low density. The evolution of these quantities as a function of the density is shown in Fig. \ref{fig:asretilde}.
In the limit of very large $k_F$, we can expand in $(a_s k_F)^{-1}$ and we obtain to leading order:
\begin{eqnarray}
\widetilde a_s(k_F)  & \simeq & 
 - \frac{9 \pi}{10 k_F} (1-\xi_0) , \label{eq:asinf}  \\
\widetilde r_e(k_F)  & \simeq &
\frac{200}{27 (\nu -1)} \frac{\eta_e}{(1- \xi_0)^2}
\frac{r_e}{\left[ 1+ \delta_e (r_e k_F)/\eta_e\right]} \label{eq:reinf}
\end{eqnarray}
It is worth mentioning that keeping these two terms in the energy, i.e. setting to zero the p-wave scattering volume, gives 
the unitary gas limit of the functional.  Results of this functional are shown by long dashed line in panel (c) of Fig. \ref{fig:func1}.

If we further take the LO in the expansion of $(r_e k _F)^{-1}$ in Eq. (\ref{eq:reinf}) we deduce for the asymptotic equation:
\begin{eqnarray}
\widetilde r_e(k_F)   & \simeq &  
\frac{200}{27 (\nu -1)} \frac{\eta^2_e}{(1- \xi_0)^2 \delta_e k_F}. \label{eq:reinf2}
\end{eqnarray}

\begin{figure}[htbp]
\includegraphics[width=1.1\linewidth]{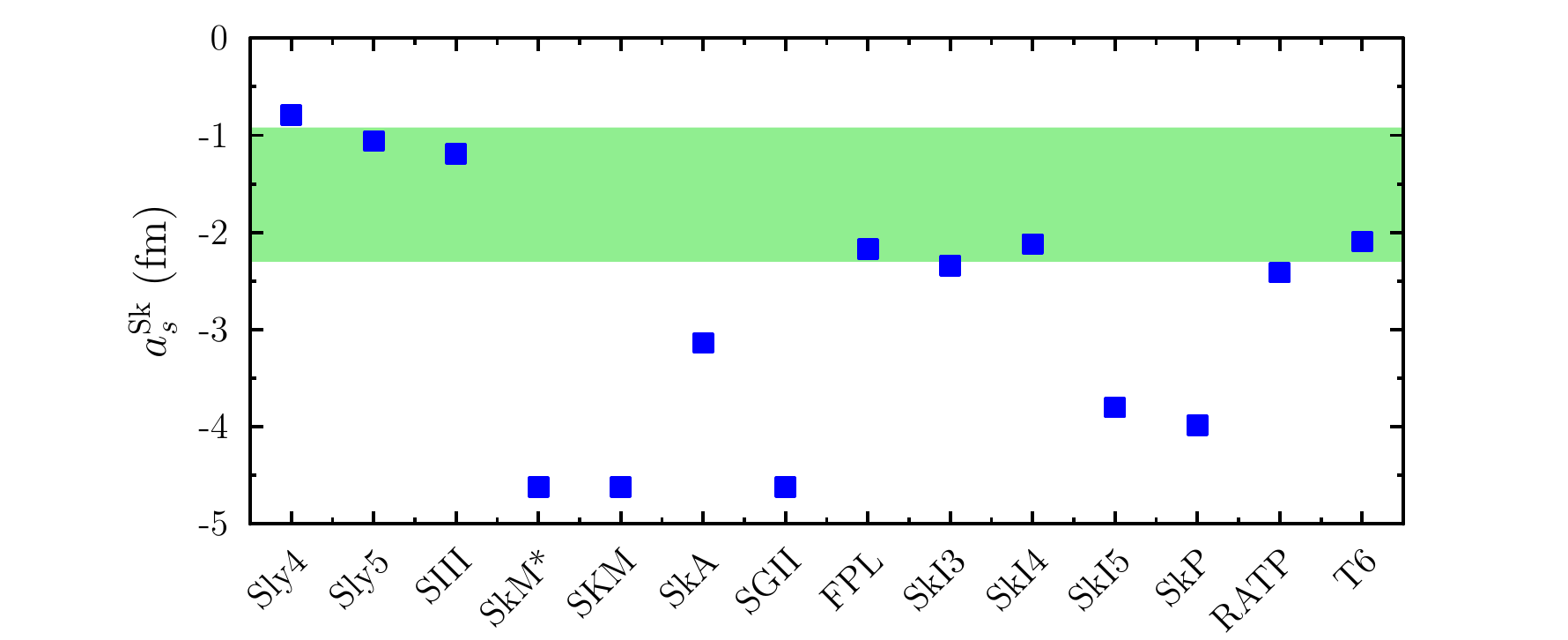} \\
\includegraphics[width=1.1\linewidth]{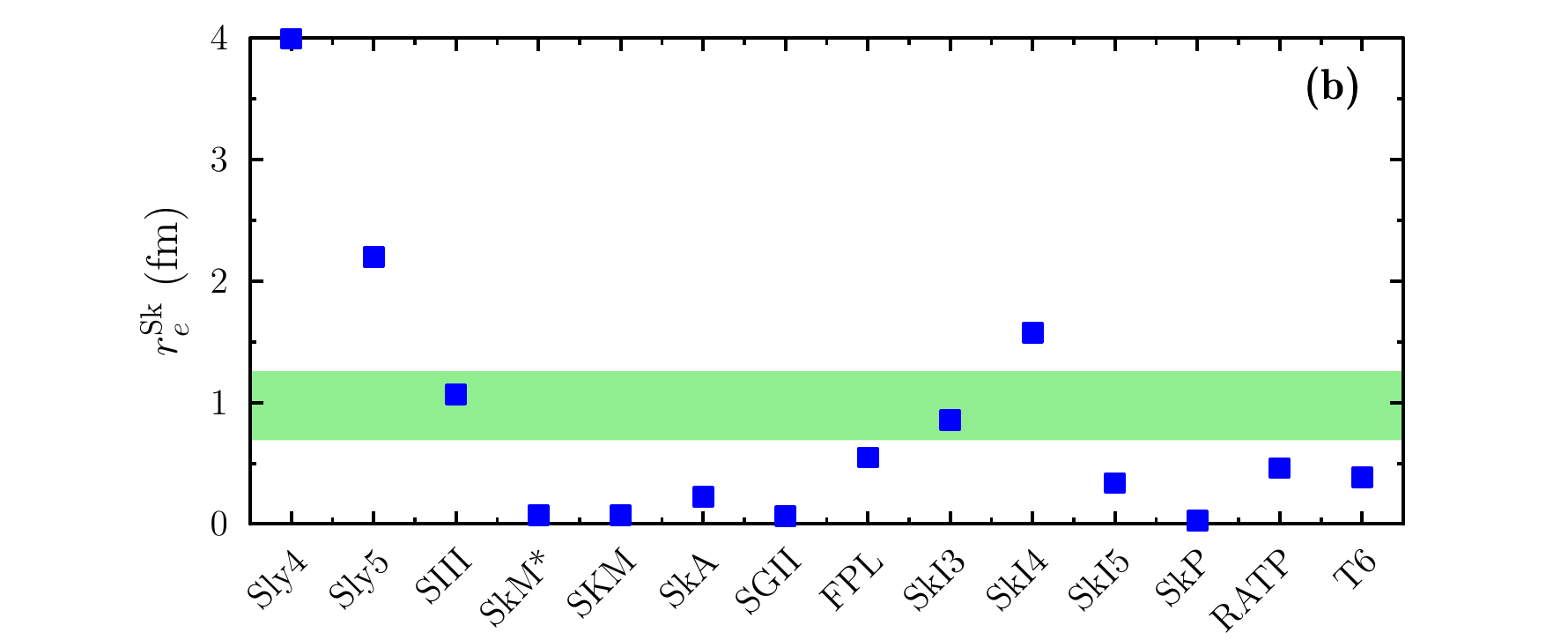} \\
\includegraphics[width=1.1\linewidth]{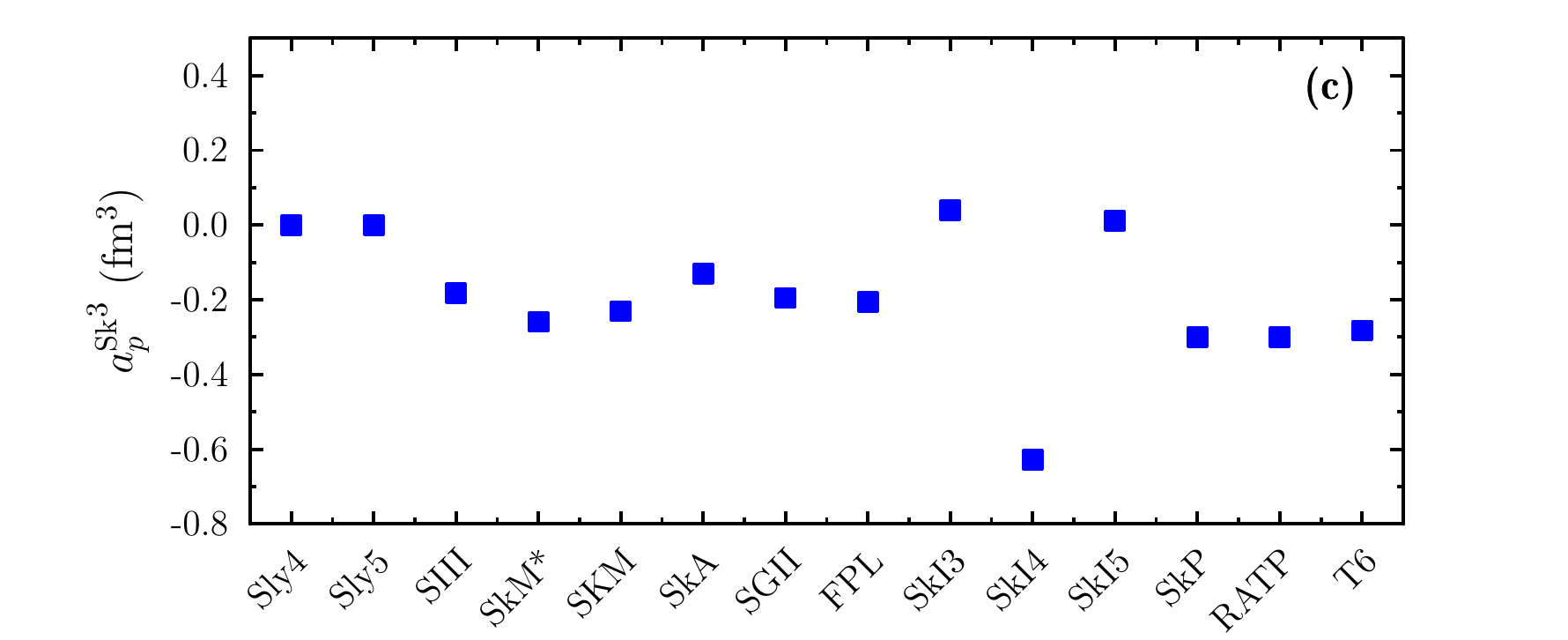}  \\
\vspace{-5mm}
\caption{Values of the quantities (a) $a^{\rm Sk}_s$, (b) $r^{\rm Sk}_e$,  for different widely used sets of Skyrme interaction parameters. 
In panels (a) and (b), the green area corresponds to the two windows of  $\tilde a_s(\rho)$ and $\tilde r_e(\rho)$ values 
given by (\ref{eq:wind}). For completeness we also give in panel (c) the equivalent to the p-wave scattering volume values, 
${a^{\rm Sk}_p}^3$, for the same set of Skyrme parameters.}
\label{fig:arskyrme} 
\end{figure}

There are several interesting conclusions one can draw from Fig. \ref{fig:asretilde}:
\begin{itemize}
  \item We can see two regimes of evolution of  $\tilde a_s(\rho)$ and $\tilde r_e(\rho)$, first at very low density, 
  they evolve very fast to much lower absolute values compared to their bare values. Then, they present a
  much smoother evolution towards higher densities. The sharp decrease is due to the strong influence of $-(a_s k_F)$ 
  terms at very low density that tends rapidly to zero due to the very large value of $a_s$. This is illustrated by comparing 
  the complete evolution (solid line) with the LO order in the expansion of $-(a_sk_F)^{-1}$ (short dashed line).
  \item Strictly speaking, the present functional has been validated up to densities $\rho < 0.01$ fm$^{-1}$. 
  The approximate expressions (\ref{eq:asinf}) and (\ref{eq:reinf}) are already very accurate. In addition, already at these densities, the effective values of the  
  s-wave scattering length and effective range are strongly reduced compare to $-18.9$ fm and 2.7 fm. 
  \item One of the most surprising conclusion one can draw from the present analysis is that the s-wave 
  scattering length has completely disappeared from the expressions (\ref{eq:asinf}) and (\ref{eq:reinf}). 
  In particular, $\tilde a_s(\rho)$ has become independent of its value in the vacuum and its value is solely determined 
  by the universal unitary gas parameters.  This gives in particular an explanation why the parameters used to reproduce 
  nuclear systems at equilibrium differ completely from those valid at low density \cite{Fur12}.  
  \item It is worth mentioning that the expression (\ref{eq:reinf2}) where both $a_s$ and $r_e$ have disappeared does not 
  reproduce the full expression while the approximate form (\ref{eq:reinf}) provides a good approximation for densities 
  $\rho \ge 0.1$ fm$^{-3}$. Therefore, the connection of $\tilde r_e$ to $r_e$ partially persists. 
   
  \item When the energy density functional becomes independent of the scale $a_s$ at higher densities, the evolution 
  of  $\tilde a_s(\rho)$ and $\tilde r_e(\rho)$ is much slower. For instance, when the density increases from $0.01$ fm$^{-3}$ to 
  $0.2$ fm$^{-3}$, that are the densities of typical relevance for nuclear systems, we have:
  \begin{eqnarray}
  \left\{ 
  \begin{array}{c}
\displaystyle -2.3 ~{\rm fm} \le \tilde a_s(\rho) \le  - 0.92 ~{\rm fm},  \\
 \\
\displaystyle +0.69 ~{\rm fm}  \le \tilde r_e (\rho) \le  +1.26 ~{\rm fm} . \\
\end{array}
\right.
\label{eq:wind}
\end{eqnarray}
These values can then be compared to the equivalent values obtained using Skyrme  functional. The equivalent values 
of the s-wave scattering length, effective range and p-wave scattering volume, denoted respectively by $a^{\rm Sk}_s$, 
$r^{Sk}_e$ and $(a^{\rm Sk}_p)^3$ can be obtained using the three equations  (\ref{eq:lec1}-\ref{eq:lec3}) where the $t_i$
and $x_i$ parameters are the standard Skyrme parameters.  Several examples obtained with different sets of Skyrme parameters 
are illustrated in Fig. \ref{fig:arskyrme}.  We see that the windows given in  (\ref{eq:wind}) are of the same order of magnitude 
compared to the Skyrme values.  We also give in panel (c) of Fig. \ref{fig:arskyrme}  the equivalent p-wave scattering volume for Skyrme 
forces. We see that this volume is often negative and does not match the value relevant at low density, i.e. $a^3_p = 0.6$ fm$^{3}$. This suggests 
that a treatment equivalent to what has been done in the s-wave should also be made for the p-wave.   
We should however keep in mind that Skyrme functionals are globally adjusted and the physical interpretation of each separate terms
as coming from a single-channel is a priori impossible, thus $(a^{\rm Sk}_p)^3$ could receive contributions from higher partial-waves (such as d-waves) as well.

\item As a side remark, one could note that, within our approach, neither $\widetilde C_0(k_F)$ nor $\widetilde C_2(k_F)$ 
are constant
parameters contrary to parameters in Skyrme functionals.  The fact that latter 
functionals work so well might stem from the slow variations observed  in the insets of Fig. \ref{fig:asretilde}.

\item The close agreement between the order of magnitude of the length (i.e., $a_s$, $r_e$) obtained in the new functional proposed here and the 
one deduced from Skyrme parameters is, to our opinion, a very interesting outcome of the present study. Indeed, since Skyrme or other 
functionals are adjusted directly on expected properties in infinite systems or on experimental observation in finite 
systems, we a priori loose track to the underlying fundamental constants directly linked to the interaction or unitary limit. The present finding 
however opens new hopes to get functionals close to the simple Skyrme ones built on first principles.   
 \end{itemize}
  
\section{Possible extension of the functional from low density to saturation density}  
\label{sec:improvement}

The functional (\ref{eq:funcunit}) seems appropriate from very low density up to $\rho\equiv 0.01$ fm$^{-3}$ (see Fig. panel (c)
and (d) of Fig \ref{fig:func1}). If we assume that the unitary gas limit is an adequate starting point (long dashed line in Fig. \ref{fig:func1}), the discrepancy observed with our new functional and the ab-initio calculations at higher densities might be due to: (i)  
effects  of higher partial waves of the two-body interaction; (ii) possible three-body interaction effects; (iii) the shape of the functional 
itself and in particular a too strong effect of $r_e$ at high density.   

Here, we explore the possibility to slightly modify the $r_e$ dependence such that the unitary and low density
limit is still properly described while better reproducing ab-initio results for $\rho > 0.01$ fm$^{-3}$. As shown in section \ref{sec:gauss}, 
the functional form is strongly guided by the re-summation of $r_e$ effect for Hartree-Fock energy obtained with a 
finite-range interaction. More precisely, we used the approximation
\begin{eqnarray}
F(x) & \simeq &  \frac{1}{1+ \frac{3}{10} x^2},   \label{eq:resumx}
\end{eqnarray}  
where $F(x)$ is given by Eq. (\ref{eq:f2}). 
The re-summed expression is designed such that the Taylor expansion up to $(\mu k_F)^2$ is the same and 
that the approximate functional remains close to the exact $F$ at larger $x$ values. Obviously, the function used for re-summation is not unique   
and alternative form can be used.  
In particular, as discussed from Fig. \ref{fig:gaus}, the convergence towards the limit 
$1 + \frac{5}{9 \pi} (a_s k_F)$ is faster in the exact case, due to the Gaussian appearing in the $F$ function given by Eq. (\ref{eq:f2}). 
Another functional with improved property and that keeps the form Eq. (\ref{eq:resumx}) as a starting point 
can be simply obtained by using
\begin{eqnarray}
F(x) & \simeq & 1 - \frac{3}{10} x^2 + \frac{9}{140} x^4 + \cdots \nonumber \\
&\simeq& \frac{1}{1+ \frac{3}{10} x^2} e^{-\alpha  x^4},\nonumber
\end{eqnarray}  
with $\alpha= 9/350$. Such generalized expression improves slightly the approximate energy evolution compared 
to the exact HF one especially at large $k_F$. Again, this illustration can only give us a guidance to modify our functional since 
we are dealing here with strongly interacting systems.

Based on this simple example, the simplest way to reduce the effect of $r_e$ in the functional (\ref{eq:funcunit})
while keeping all the nice properties unchanged is to multiply the second term in Eq. (\ref{eq:funcunit}) by $e^{-c(a_s/r_e) (r_e k_F)^4}$. Here $c(a_s/r_e)$ is a new parameter that should a priori be fixed with appropriate arguments. For instance, in the example 
given above, using the fact that $x = (a_s r_e k_F^2)$, we deduce $c(a_s/r_e) \simeq 0.026 (a_s/r_e)^2$.  
\begin{figure}[htbp]
\includegraphics[width=\linewidth]{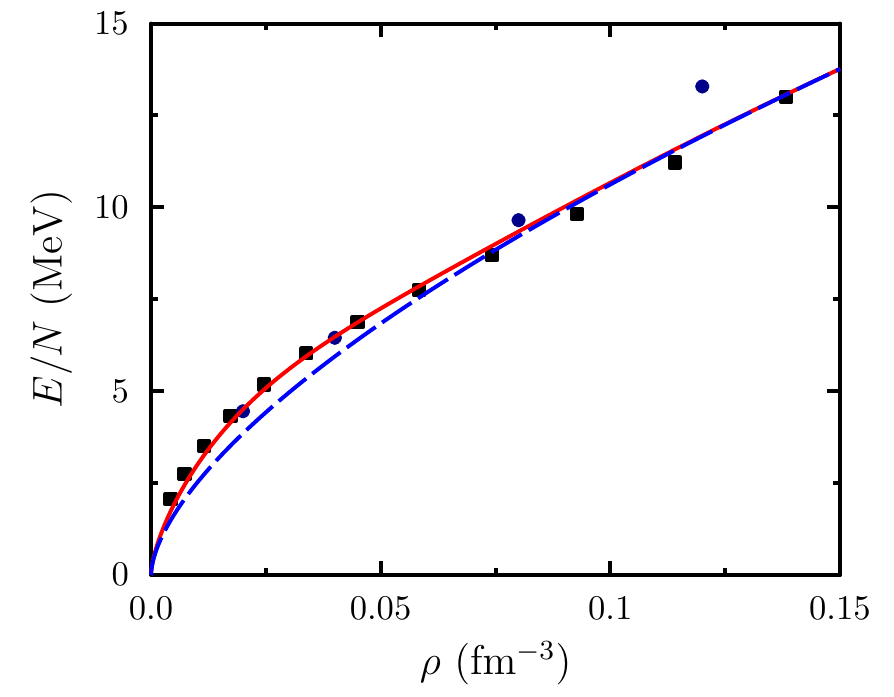}
\vspace{-5mm}
\caption{ (color online)  Same as panel (d) of Fig. \ref{fig:func1}. The red solid line represents the result of the functional with the extra 
factor $\displaystyle e^{-c (r_e k_F)^4}$ in the $r_e$ dependence. }
\label{fig:damp} 
\end{figure}

There are arguments to assume that the $c$ parameter should be independent of $a_s$ for large densities.
The first one is very practical. Indeed, if we suppose that this parameter is proportional to $a^2_s$, it will cancel the $r_e$ dependence
at unitarity and we cannot impose anymore  the constraint (\ref{eq:etae}).   The second argument, more fundamental, stems from our 
previous conclusion that at densities above 0.01 fm$^{-3}$, the scales $a_s$ become irrelevant.  At this stage, we simply assume that 
$c$ is a constant that is independent of $a_s$ and should a priori be obtained from the unitary gas properties.  Its determination would require 
to have an extra term in the expansion (\ref{eq:etae}). Since we do not have it, and as a proof of principle, we simply adjust this term to 
reproduce the ab-initio result of Refs. \cite{Fri81} up to density $\rho=0.16$ fm$^{-3}$. Doing so, we leave our original strategy to have no free parameters, hoping that future progress on unitary gas with effective range will justify the retained value of $c$.  In Fig. \ref{fig:damp}
we show the result obtained with $c=0.02$, keeping all parameters of the functional equals to their previous values given in the set of equations (\ref{eq:param}). We see in particular, as expected, that the low--density regime is still perfectly reproduced while a much better  
agreement is obtained at high density. 
 
 \section{Summary and discussion}
 
 In the present article, following the work of Ref. \cite{Lac16}, we further discuss the possibility to develop 
 nuclear DFT using the unitary regime as a starting point.  One of the clear advantage of the present approach 
 is that the functional has no free parameters and depends explicitly on the physical low--energy constant as well 
 as on the universal parameters describing Fermi systems at unitarity. Several aspects of the functional proposed 
 in Ref. \cite{Lac16} are clarified. We show that the good matching of the functional with exact QMC approach 
illustrates that nuclear systems can be treated perturbatively in $-(a_s k_F)^{-1}$ with respect to the unitary gas.

An important advantage of the present approach compared to other functionals based on bare LEC like Lee-Yang EDF, is that it can 
be applied for densities that starts to be of relevance for nuclear systems.  By defining density--dependent scattering length and effective 
range, we analyze how these quantities makes a transition from the very low density regime to higher 
densities. We show that the relevant scales are strongly renormalized at very low density. This rapid evolution stems from 
the anomalously large s-wave scattering length in nuclear systems. After this rapid evolution, the relevant scales stabilize and slowly
evolve. An important conclusion we draw is that the smooth evolution is completely dominated by the universal constant at unitarity for
the s-wave. In particular, this scale becomes independent of its bare value. The situation is slightly different for the 
effective range.  Its evolution depends on both 
the unitary regime and its bare limit $r_e$. Since the s-wave scattering length 
is the only scales that is anomalous large, we do anticipate 
that the behavior observed for the effective range should be the same for other scales in other channels.  

One of the interesting byproduct of the present work is that it gives some preliminary steps towards explaining why 
simple functionals, like Skyrme functionals, can be so successful while the apparent associated scales completely 
differs from the physical ones at low densities.  At the heart of the reasoning is that part of the scales important in nuclear physics 
are not the one at this regime but the one at unitarity and therefore are independent on the underlying interaction.
  
A key aspect of the present work is the useful recent progress made on nuclear interactions, the precise study of systems 
at unitarity and the possibility to obtain exact solutions for nuclear systems and/or cold atoms in different regimes of s-wave 
scattering length. While the strategy we used here to design a nuclear DFT without any adjustable parameter is unambiguous, we are still 
far from having a predictive functional for densities up to twice the saturation density. Indeed, up to now we concentrated our attention 
to the neutron matter and incorporated essentially the s- and p-wave channels that are the dominant at low density. To further progress, other states of matter with various spin/isospin contents should be considered 
together with  higher orders partial waves. 

In the present exploratory study, the functional is solely designed for neutron systems. This restricts the range of applicability to neutron matter. One could also envisage the description of neutron droplets using for instance the local density approximation to treat finite systems. Work is in progress along this line. A great challenge to render the approach more versatile would be to extend it to nuclear matter and more generally to asymmetric matter. It should be noted that this project should 
be made back-to-back with progress in ab-initio calculations, like QMC theory, to obtain exact benchmark calculations of increasing complexity including effect beyond direct two-body interactions.


\begin{acknowledgments}
The authors thanks A. Gezerlis for useful discussion at different stage of the work and for providing a crosscheck 
of the p-wave scattering volume for the AV4 interaction. This project has received funding from
the European Unions Horizon 2020 research and innovation
program under grant agreement No. 654002.
\end{acknowledgments}

\end{document}